\documentclass[10pt]{article}

\usepackage{amsfonts,amssymb,amsmath}
\usepackage{graphicx}
\DeclareGraphicsExtensions{.pdf,.png,.jpg}
\begin{document}

\title{\bf High--N00N State Generation: N00N State Output of Floquet Engineering }
\author{ Yusef Maleki
\thanks{E-mail: maleki@physics.tamu.edu}
\\{\small Institute of Quantum Science and Engineering, Department of Physics and Astronomy,}
\\{\small College Station, Texas 77843-4242, USA}}

 \maketitle

Here, we review some quantum architectures designed for the engineering of the N00N state, a bipartite maximally entangled state crucial in quantum metrology applications.  The fundamental concept underlying these schemes is the transformation of the initial state $|N\rangle
\otimes |0\rangle$ to the N00N state $\frac{1}{\sqrt{2}} (|N\rangle
\otimes|0\rangle +|0\rangle \otimes|N\rangle)$, where $|N\rangle$ and $|0\rangle$ are the Fock states with $N$ and $0$ excitations. We show that this state can be generated as a superposition of modes of quantum light, a combination of light and motion, or a superposition of two spin ensembles.
The approach discussed here can generate mesoscopic and macroscopic entangled states, such as entangled coherent and squeezed states, as well. We show that a large class of  maximally entangled states can be achieved in such an architecture.
The extension of these state engineering methods to the multi-mode setting is also discussed.

\maketitle

\newpage

\section{Introduction}

Estimating unknown parameters stands as a pivotal task across diverse scientific disciplines, playing an important role in technological advancements  and scientific discoveries \cite{Helstrom, Holevo, Caves, Braunstein1, Giovannetti, Agarwal,Kashami2024,Abbott}. The task of attaining the highest possible accuracy in parameter estimation is a fundamental objective within the field of metrology \cite{Maleki-2022, Bevington, Sanders1, Boto}. One effective strategy for enhancing estimation accuracy involves the repetition of experiments and the collection of additional information on the parameter. Specifically, utilizing $N$ independent resources for measuring the parameter $\varphi$, the optimal sensitivity of the parameter is determined by the central limit theorem, scaling as $\Delta\varphi\propto1/{\sqrt{N}}$. This phenomenon is commonly referred to as the shot-noise limit or standard quantum limit \cite{Lee, Demkowicz, Maleki1}.

In the realm of parameter estimation, quantum physics introduces an intriguing capability that remains beyond the reach of classical methodologies \cite{ Ono, maleki2018generating, Dowling, Cappellaro}. 
In fact, quantum physics has the ability to push the boundaries of precision, enabling measurements that were once deemed impossible within classical frameworks  \cite{Maleki2023, Afek, Maleki2021,Maleki2021Metrology}. Quantum-enhanced precision opens the door to unprecedented accuracy, revolutionizing fields of science and technology where extreme accuracy of the measurement is the key \cite{Holland93, Zhou2018}. The exploration of quantum capabilities in parameter estimation has become a focal point of research, promising breakthroughs that can change our understanding of measurement precision.

Therefore, quantum resources offer the potential for a substantial enhancement in precision, surpassing classical limits and approaching the fundamental bounds dictated by quantum mechanics. This quantum-enhanced precision can go beyond the classical limit and reach the Heisenberg Limit (HL), where the estimation error $\Delta\varphi$ scales as $1/{N}$ for a given number of the resource $N$ \cite{Xiang}.
The noteworthy quantum state that can achieve this limit is the N00N state, represented as $\frac{1}{\sqrt{2}} (|N\rangle \otimes |0\rangle + |0\rangle \otimes |N\rangle)$ . This state, renowned for its potential to fulfill Heisenberg limit precision, holds immense promise for quantum-enhanced metrology \cite{Xiang}.

At the heart of this quantum advantage provided by the N00N state is the quantum entanglement manifested by such a state \cite{Xiang,Birrittella}. Quantum entanglement, defying classical intuition, showcases the unique properties of quantum systems, which can serve as the resource for various quantum applications ranging from quantum computing to quantum communication and beyond \cite{Khashami2013, Maleki2016-1}.
In recent years, there has been a considerable effort to utilize entangled states for quantum sensing and metrology.

 However, the realization of the N00N state poses a considerable challenge, restricted by difficulties in generation and susceptibility to decoherence \cite{Xiang}. Of course, such difficulty is not limited to the N00N state generation and remains a challenge within quantum science, in general, \cite{Lidar1, Lidar2,   Maleki2018JOSA, Maleki2019OE, Maniscalco, Basit}.
The difficulty of generating entangled states in quantum systems is a challenge that researchers are actively addressing. The inherent difficulty in achieving the desired entangled states reflects broader challenges in manipulating quantum states for practical applications. Overcoming these challenges is essential for harnessing the full potential of quantum-enhanced precision in parameter estimation scenarios.

This review considers intricate  quantum architectures designed to engineer entangled states, focusing on the N00N state generation. 
We demonstrate the generation of this state through a superposition of quantum light modes, a combination of light and motion modes, or a combination of two spin ensemble states. The method elucidated in this discussion is capable of producing mesoscopic and macroscopic entangled states, including entangled coherent and squeezed states. Hence, we show that a broad spectrum of maximally entangled states can be realized within this architectural framework.
The exploration of these state engineering techniques is extended  to the multi-mode setting and is also thoroughly examined in our discussion.

\section{Quantum Circuit for N00N state generation}

\begin{figure}
\centering
\includegraphics[width=10 cm]{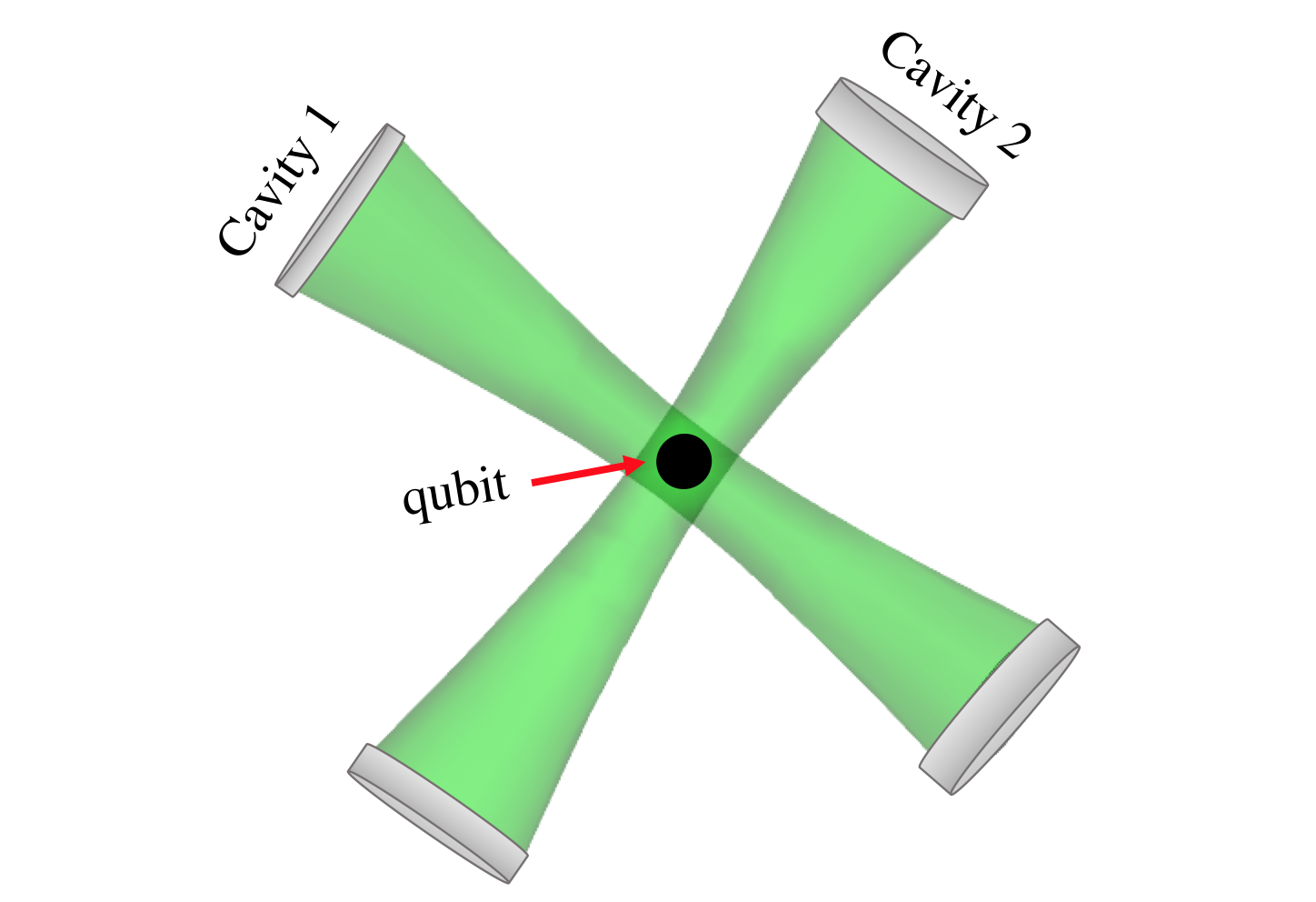}
\caption{An scheme of the generating N00N state. The system consists of two cavities interacting with a qubit. The qubit dispersively interacts with the cavities.}
\label{fig:1}
\end{figure}

To generate the N00N state, we suppose that we have an interaction Hamiltonian of two cavity systems with a two-level qubit given as \cite{maleki2018generating, maleki2019high}

\begin{equation}\label{N00Nham1}
H\textsubscript{eff}= i \kappa(a_1a^{\dagger}_2- a_2a^{\dagger}_1)\sigma_z,
\end{equation}

in which the  boson annihilation and
creation operators of the first and the second modes are $a_1, a_1^\dag$ and $a_2, a_2^\dag$, respectively. These operators can be associated with any harmonic oscillator system, such as photons in the cavity \cite{Maleki2021OC, Maleki2017}. The qubit operators in this equation, $\sigma_z= |1\rangle\langle 1|-|0\rangle\langle 0|$,  $ \sigma_+= |1\rangle\langle 0| $ and $ \sigma_-= |0\rangle\langle 1| $  are the Pauli operators, that obey su(2) Lei algebra. Therefore, this Hamiltonian can be considered as  a two-level system (qubit) interacting with two bosonic modes, such as two modes of the light fields.

As was mentioned earlier, central to our approach for the high-N00N state generation is the derivation of the effective Hamiltonian in Eq. (\ref{N00Nham1}). For simplicity, one can consider two cavity fields interacting with a single qubit as in Fig. \ref{fig:1}. We will provide a detailed derivation of this effective Hamiltonian later; however, it is stimulating to see how such a Hamiltonian results in a generation of N00N state, starting from $N$ photons in one of the resonators. In  fact, we are going to show that, starting from this interaction Hamiltonian, one can generate the N00N state in the two modes. To this aim, we consider the qubit to be in the logical superposition state $
\frac{1}{\sqrt{2}}(|0\rangle+|1\rangle)$. The first and the second field modes are prepared in the $|0\rangle_1$ and $|N\rangle_{2}$ Fock states, respectively. Therefore, the initial state of the composed system, including the atom and the cavities, is thus given by 
$$
|\psi(0)\rangle= \frac{1}{\sqrt{2}}(|0\rangle+|1\rangle)|0\rangle_{1}
|N\rangle_{2}.
$$
Starting from this initial state, we consider the time evolution of the entire system with the effective Hamiltonian above. This implies that the time evolution operator 
$$
U(t)=e^{-{iH\textsubscript{eff} \Delta t}/{\hbar}},
$$
 should be applied to this initial state, from which we find \cite{maleki2018generating}
\begin{equation}\label{}
|\psi(t)\rangle= \frac{1}{\sqrt{2}}(e^{-{iH\textsubscript{eff} \Delta t}/{\hbar}}|0,N\rangle|0\rangle+e^{-{iH\textsubscript{eff} \Delta t}/{\hbar}}|0,N\rangle|1\rangle),
\end{equation}
where $|0,N\rangle\equiv|0\rangle_{1}
|N\rangle_{2}$. 

We let the system evolve for the time interval  $ \Delta t_1=\pi/(4 \Omega).$ This entails the time evolved state $|\psi(t)\rangle$ to degenerate into  

\begin{equation}\label{}
|\psi\rangle_1= \frac{1}{\sqrt{2^N}} \sum_{k=1}^N \left(
  \begin{matrix}
    N  \\
   k
  \end{matrix}
  \right)^\frac{1}{2}
|k,N-k\rangle(|0\rangle+(-1)^k|1\rangle).
\end{equation}
We note that in this state, the two fields and the qubit become entangled. The measurement of the first resonator in the state $|k\rangle$ projects the qubit in two orthogonal superposition of the computational basis. Namely, detecting the first resonator in the state with an even $k$ projects the qubit into the state $\frac{1}{\sqrt{2}} (|0\rangle+|1\rangle)$, and with odd $k$ projects the qubit into the state $\frac{1}{\sqrt{2}} (|0\rangle-|1\rangle)$. These two basis are orthogonal. Alternatively, detecting the qubit in these two orthogonal basis provides information about the $k$. Similar argument can be made for the second mode.

Next, we apply the Hadamard gate to this state. The Hadamard gate is given by a $2\times 2$ matrix acting on the Hilbert space of a single qubit which is given by
\begin{equation}\label{}
\mathcal{H} =  \frac{1}{\sqrt{2}}\left( \begin{matrix} 1&1\\ 1&-1 \end{matrix} \right).
\end{equation}
The role of this gate is to rotate the basis of the Hilbert space of the qubit. Hence,  applying this gate translates into a local transformation of the qubit basis such that
$|0\rangle\rightarrow \frac{1}{\sqrt{2}}(|0\rangle+|1\rangle)$, and $|1\rangle\rightarrow \frac{1}{\sqrt{2}}(|0\rangle-|1\rangle)$. Once the qubit is subjected to the Hadamard gate, we let the system to undergo the time evolution again within the time interval  $ \Delta t_2=\pi/(4 \Omega)$. With this time evolution the quantum state of the system reduces to   $|\psi\rangle_2$, which is explicitly given by
\begin{equation}
|\psi\rangle_2= \frac{1}{\sqrt{2^N}} \sum_{k=1}^N \left(
  \begin{matrix}
    N  \\
   k
  \end{matrix}
  \right)^\frac{1}{2} [1+(-1)^k]
|k,N-k\rangle|0\rangle
+\frac{1}{\sqrt{2^N}} \sum_{k=1}^N \left(
  \begin{matrix}
    N  \\
   k
  \end{matrix}
  \right)^\frac{1}{2} [1-(-1)^k]
|k,N-k\rangle|1\rangle.
\end{equation}\label{}

Next, we measure the qubit in its computational basis, which projects the field modes into the superposition of the number states. If  the outcome of the measurement performed on the qubit gives $|0\rangle$, the wave function of the system collapses to 
\begin{equation}\label{}
|\psi\rangle_3= \frac{1}{\sqrt{2^N}} \sum_{k=1}^N \left(
  \begin{matrix}
    N  \\
   k
  \end{matrix}
  \right)^\frac{1}{2} [1+(-1)^k]
|k,N-k\rangle|0\rangle.
\end{equation}
Clearly, the qubit becomes disentangled from the rest of the system due to the wave function collapse.

Alternatively, if the outcome of the measurement  gives $|1\rangle$, then the wave function of the system collapses to 
\begin{equation}\label{}
|\psi\rangle_3= \frac{1}{\sqrt{2^N}} \sum_{k=1}^N \left(
  \begin{matrix}
    N  \\
   k
  \end{matrix}
  \right)^\frac{1}{2} [1-(-1)^k]
|k,N-k\rangle|1\rangle
\end{equation}
Once we measure the state of the qubit, regardless of the outcome of the measurement, we apply the time evolution operator one more time on the collapsed wave function of the system. First, let us assume the outcome of the measurement to be $|0\rangle$. Applying the time evolution operator on the corresponding collapsed wave function,  for a time interval $\Delta t=\pi/(4 \Omega)$ following the measurement step results in the N00N state 
\begin{equation}\label{}
|\textrm{N00N} \rangle= \frac{1}{\sqrt{2}} ((-1)^N|N\rangle_{1}
\otimes|0\rangle_{2} +|0\rangle_{1} \otimes|N\rangle_{2}).
\end{equation}

Alternatively, if the  outcome of the measurement of the qubit gives $|1\rangle$, then, once we  let the system to evolve for a time interval $3\pi/(4 \Omega)$, one arrives at the N00N state of the form 
\begin{equation}\label{}
|\textrm{N00N} \rangle= \frac{1}{\sqrt{2}} (|N\rangle_{1}
\otimes|0\rangle_{2} -|0\rangle_{1} \otimes|N\rangle_{2}).
\end{equation}

It is worth noting that, regardless of what the outcome of the measurement of the qubit state is, the N00N state can be obtained at the final step of this protocol. This implies the fact that considering the ideal scenario where no decoherence is involved and the measurement is perfect, we arrive at the  $100\%$ fidelity of N00N state generation from $|\psi\rangle_2$. Therefore, the N00N state generated with this protocol is deterministic. If the measurement yields $|0\rangle$, it takes the system less time to evolve to the N00N state. 

\begin{figure}
\centering
\includegraphics[width=14 cm]{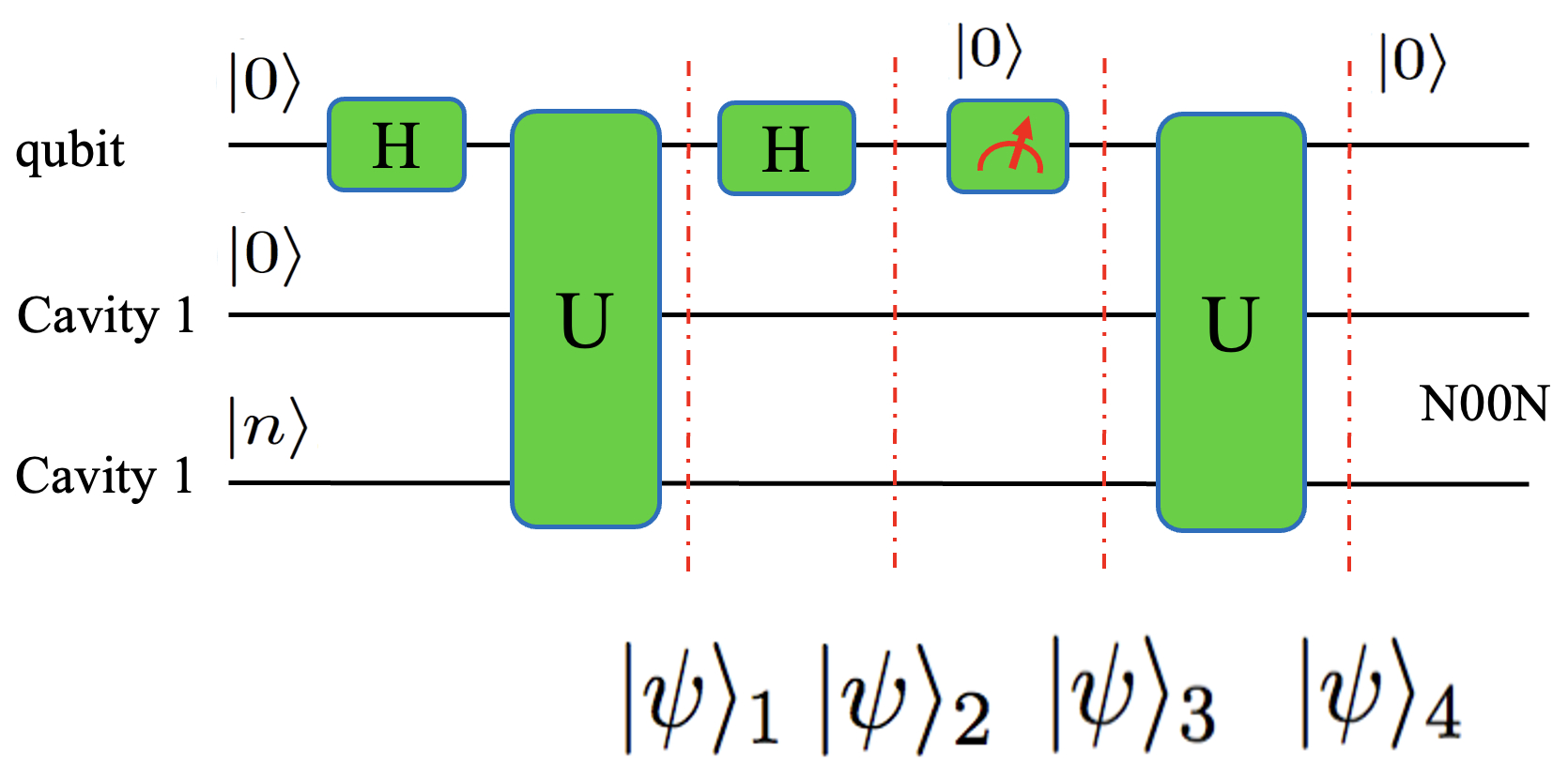}
\caption{
N00N state generation circuit: U represents the time evolution of a quantum state, governed by the time-evolution operator with the effective Hamiltonian $H\textsubscript{eff}$. $\mathcal{H}$ is the Hadamard gate acting locally on the qubit. For the time interval $\Delta t_1=\pi/(4 \Omega)$, the system evolves from its  initial state to $|\psi\rangle_1$. The Hadamard gate yields the state $|\psi\rangle_2$. Once a measurement on the state of the flux qubit is performed, the total wave function of the systems collapses to $|\psi\rangle_3$. finally, time evolution $U(t)$ provides the desired N00N state.}
\label{fig:Fig-M-2}
\end{figure}

It is interesting to note that, once the Fock state $|N\rangle$ is provided, the number of steps that are needed to arrive at the N00N state is independent of the number of photons $N$. Most of the N00N state generation schemes start from $N=1$ N00N state and increase the $N$ in several steps. In these schemes, the number of the operations is directly related to the number $N$; however, in the architecture given in this chapter, this may not be the case. This is  important as it can provide a better control of decoherence. In fact, once we prepare a high-fidelity Fock state $|N\rangle$, in one of the modes, without getting the rest of the system involved in the process, we can translate that into a N00N state in just a few operational steps. Thus, this scheme is more robust to decoherence. On the other hand, contrary to most of the proposed protocols, this scheme is simpler and requires fewer devices. We only need a single qubit which interacts with two modes.
The architecture described above is schematically given in Fig. (\ref{fig:Fig-M-2}).

\section{Derivation of the effective Hamiltonian}
\subsection{Modulating the coupling between the qubit and the resonators:}

We consider a setup of two resonators interacting with a same two-level atom, where the Hamiltonian can be written as a summation of the free part and the interaction part $H=H_0+H_1$ \cite{maleki2019high}. We assume that the coupling  parameter of the cavity fields and the qubit is periodically modulated. Therefore, the free part of the Hamiltonian is
$$
H_0= \hbar\dfrac{\omega_0}{2}\sigma_z+ \ \hbar \omega_1 a_1^{\dagger} a_1+ \ \hbar \omega_2 a_2^{\dagger} a_2,
$$
and the interaction part reads
 $$
 H_1= 2 \hbar g_0 \sum_{j=0}^2\ \cos(\upsilon_j t+\varphi_j)(\sigma^{+}a_j+a^{\dagger}_j\sigma^{-}).
 $$

Here, $\omega_0$ is the frequency of the two-level energy splitting  between its ground and exited  levels. $\omega_1$ and $\omega_2$ are the frequencies of the  first and the second cavities, respectively. $g_0$ is the coupling strength of the resonators and the qubit. $\upsilon_j$ is the modulating frequency and $\varphi_j$ is the modulating phase  of the resonator $j$, with $j=1,2$.
We consider the resonant case where $\omega_1=\omega_2=\omega$ and 
take $\hbar=1$, for simplicity. Now, with this Hamiltonian, we are going to extract the effective Hamiltonian of the N00N state that we already discussed. To this aim we define the unitary evolution $U_0(t)  = \exp[-iH_0 t]$, using the free part of the
the Hamiltonian. Consequently,  in the rotating frame  the interaction Hamiltonian can be given as 
$$
 H_I= U_0(t) H U^{-1}_0(t)-H_0.
 $$
The explicit form of the interaction Hamiltonian, thus, reads
$$
 H_I=\delta  \sigma_z/2+ 2  g_0 \sum_{j=1}^2\ \cos(\upsilon_j t+\varphi_j)
 (\sigma^{+}a_j+a^{\dagger}_j\sigma^{-}), 
 $$
 where we have  $\delta=\omega_0-\omega$. If we express the $\cos(\upsilon_j t+\varphi_j)$  term as an exponential form, the Hamiltonian equivalently can be expressed as 
\begin{equation*}
H_I=\delta  \sigma_z/2+ g_0 \sum_{j=1}^2\ [(\sigma^{+}a_j+a^{\dagger}_j\sigma^{-})e^{i\varphi_j}]e^{i\upsilon_j t}
+[(\sigma^{+}a_j+a^{\dagger}_j\sigma^{-})e^{-i\varphi_j}]e^{-i\upsilon_j t}.
\end{equation*}
This Hamiltonian has two parts which are related to the time-dependent terms $e^{i\upsilon_j t}$ and $e^{-i\upsilon_j t}$, respectively. Therefore, we define the  operator $h_j$ such that
$$
h_j= g_0 (\sigma^{+}a_j+a^{\dagger}_j\sigma^{-})e^{i\varphi_j}.
$$
With this definition, the Hamiltonian reduces to 
$$
H_I= \delta  \sigma_z/2+ \sum_{j=1}^2\ h^{\dagger}_j e^{i\upsilon_j t}+ h_j e^{-i\upsilon_j t}.
$$
This is a Floquet Hamiltonian \cite{Wang,Wang4, Goldman,Jotzu}, which will enable us to further reduce the Hamiltonian to attain the effective Hamiltonian of the desire.

Without loss of generality, we consider the same modulation frequencies $\upsilon_j=\upsilon$ in the following. Also we note that $h_j$ commutes with its Hermitian conjugate part, such that $ [h_j,h^{\dagger}_j]=0$. With these considerations, the effective Hamiltonian reduces to
$$
H\textsubscript{eff}= \delta  \sigma_z/2+i \dfrac{2g^2_0}{\upsilon} \sin(\varphi_1-\varphi_2)(a_1a^{\dagger}_2- a_2a^{\dagger}_1)\sigma_z.
$$
Thus, the coupling strength of the Hamiltonian can be directly controlled by the modulation phase factor of each cavity. To maximize the   coupling strength in the Hamiltonian we take $\varphi_1-\varphi_2={\pi}/{2}$, and define $\kappa=\dfrac{2g^2_0}{\upsilon} $, where the Hamiltonian can be expressed as 
\begin{equation}
H\textsubscript{eff}=\delta  \sigma_z/2+ i \kappa\sigma_z(a_1a^{\dagger}_2- a_2a^{\dagger}_1).
\end{equation}
Taking the resonance case $\delta=0$, we attain the desired Hamiltonian of the system as 
\begin{equation}
H\textsubscript{eff}= i \kappa\sigma_z(a_1a^{\dagger}_2- a_2a^{\dagger}_1).
\end{equation}

It is remarkable that, the term $\sin(\varphi_1-\varphi_2)$ controls the coupling strength of the scheme, here. Therefore, it is highly important to have precise control of the modulation of the system in practical scenarios. It is clear that when the modulation phases are identical, the term $\sin(\varphi_1-\varphi_2)$ becomes zero, and no effective Hamiltonian can be attained in this case.

The maximum coupling is determined by the term $\kappa=\dfrac{2g^2_0}{\upsilon} $. Therefore, even though the modulation frequency $\upsilon$, must be larger enough compared to the parameter $g_0$, however, if we choose it to be very large, the coupling constant of the effective Hamiltonian may be small. Hence, we need to be careful to satisfy the strong coupling regime when choosing the modulation frequency. Specifically, if we take $\upsilon=2 g_0$, the coupling strength of about $\kappa=g_0/2$, can be achieved.

\subsection{Modulating the frequencies of the resonators:}

We showed that the effective Hamiltonian of our system can be realized in the setup of two resonators interacting with the same two-level qubit by a specific modulation of the coupling of the system. Now, we are going to show that such a Hamiltonian can be achieved with a slightly different approach. In this case, we are going to consider the same setup of two resonators interacting with the same qubit, where, alternatively the resonators are modulated with the frequencies given by \cite{Maleki2023-1, Maleki2022-1}
$$
\nu_j(t)=\nu+ \Delta \sin(\nu_dt-\varphi_j).
$$ 
In this scenario, the frequencies of the resonators are modulated, with the same frequencies but different phases. The frequency of the qubit is constant and no modulation is applied on that.
Therefore, we can express the Hamiltonian of the system as 
 \begin{align}
 H= \hbar\dfrac{\omega_0}{2}\sigma_z+  \hbar \nu_1(t) a_1^{\dagger} a_1+ \hbar \nu_2(t) a_2^{\dagger} a_2
 + \hbar g_\nu \ (\sigma^{+}a_1+a_1^{\dagger}\sigma^{-}+\sigma^{+}a_2+a_2^{\dagger}\sigma^{-}).
\end{align}

Here, $g_\nu$ is the coupling constant of the atom to the resonators. Also, $ \nu_1(t)$ and $ \nu_2(t)$ are the frequencies of the resonators that are time-dependent  through the modulation term above. From this Hamiltonian one can derive 
the interaction Hamiltonian of the system. Thus, in the framework of the rotating wave approximate when $\omega_0=\nu$ the total Hamiltonian of the system above can be simplified as
$$
H_I=\hbar g_\nu\sigma^+(\hat{a}_1 e^{i\zeta\cos(\nu_dt-\varphi_1)}+\hat{a}_2 e^{i\zeta\cos(\nu_dt-\varphi_2)})+h.c.,
$$
where $\zeta=\Delta/\nu_d$.

This Hamiltonian can be reduced to a Floquet Hamiltonian by using the identity \cite{Wang,Wang4}
$$
e^{i\zeta\cos(\nu_d t+\varphi_j)}=\sum_{n=-\infty}^\infty J_{n}(\zeta) e^{in(\nu_d t+\varphi_j)},
$$
 where $J_{n}(\zeta)$ is the $n$th-order Bessel function of the first kind. Inserting this in the Hamiltonian above enables us to write the Hamiltonian in the form of a Floquet Hamiltonian which can be expressed as
$$H_I=H_{0}+ \sum_{n=1}^\infty H_n e^{in\nu_d t},
$$
where the time-independent part of the Hamiltonian is expressed in terms of the $0$th-order Bessel function of the first kind in the above expansion which reads
$$
H_{0}=\hbar g J_{0}(\zeta)(\sigma^+(\hat{a}_1+\hat{a}_2) +(\hat{a}_1^\dag+\hat{a}_2^\dag) \sigma^-)
$$
and the time-dependent part of the Hamiltonian is expressed in terms of the $n$th-order Bessel function where we can write 
\begin{align*}
H_n=\hbar g  i^n J_{n}(\zeta)[(\sigma^+\hat{a}_1 +(-1)^n \hat{a}_1^\dag\sigma^-)e^{in\varphi_1}
+(\sigma^+\hat{a}_2 +(-1)^n \hat{a}_2^\dag\sigma^-)e^{in\varphi_2}].
\end{align*} 
The interaction Hamiltonian of this form can be reduced to the effective Hamiltonian, for which we have \cite{Wang,Wang4, Goldman,Jotzu}
$$
H\textsubscript{eff}=H_{0}+ \sum_{n=1}^\infty[H_n,H_{-n}]/({n\hbar \nu}),
$$
 Therefore, with a straightforward calculation, the Hamiltonian can be written as 
$$
H\textsubscript{eff}=\hbar g J_{0}(\zeta)(\sigma^+(\hat{a}_1+\hat{a}_2) +h.c.)+i\hbar \Omega(\hat{a}_1^\dag \hat{a}_2 - \hat{a}_1 \hat{a}_2^\dag)\sigma_z.
$$
In this equation, $\Omega= g^2\chi/\nu$ is the coupling coefficient which depends on $g^2$ and $\nu$ similar to the previous scheme. In this relation the parameter $\chi$ is determined through
$$
\chi=\sum_{n=1}^\infty 2 J_{n}(\zeta)^2\sin(n(\varphi_1-\varphi_2))/n.
$$
It is evident that when the modulation phases are identical, this parameter becomes zero and we cannot attain the desired Hamiltonian, hence the modulation phases play a crucial role in this scenario. To improve the coupling of the system, we need to optimize the parameter $\chi$, which provides a direct strategy to control the overall coupling of the system.

We specifically choose $\zeta=2.40$ ($J_{0}(2.40)=0$) and $\varphi_2\neq\varphi_1$ to derive the effective  Hamiltonian \cite{maleki2019high,maleki2018generating}
\begin{equation}\label{hamilton}
H\textsubscript{eff}=i\hbar \Omega(\hat{a}_1^\dag \hat{a}_2 - \hat{a}_1 \hat{a}_2^\dag)\sigma_z
\end{equation}
To maximize the effective coupling coefficient of the Hamiltonian $\Omega$, one can maximize $\chi$  by controlling the phase difference  $\varphi_1-\varphi_2$. If we choose $\varphi_1-\varphi_2=\pi/3$, we will obtain $\chi\approx 0.628$.

The effective Hamiltonian can be realized in different quantum platforms.
In particular, one can couple two superconductor resonators to the same qubit \cite{maleki2019high,maleki2018generating}. In this case, the coupling strength between the  resonators  and the qubit can be tens of MHz \cite{Kubo2010}. With such a strategy, one is able to strongly couple  the resonators to the qubit and obtain the desired Hamiltonian. Therefore, in such a setup, one can effectively realize N00N state shared within the two resonators following the quantum circuit which was already discussed in a great details. As was already noted such an strategy provides a robust platform for realizing N00N states, with a high number of photons. Since Fock state with 15 microwave photons has already been generated in the lab, and generation of the Fock states with about 20 photons seems quite feasible, the platform described here can be useful for breaking the existing limitations of the N00N state generation.

Even though generation of such a N00N state with the method that has already investigated above, seems quite pleasing, we show that the main Hamiltonian can be realized in spin systems as well. In particular, we are interested in generation of the N00N state in nitrogen vacancy centers in diamond. This can provide many advantages considering the long coherence time of the diamonds, which enables to overcome the decoherence problem of the N00N state generation as the main challenge of this state. Furthermore, since NVEs are very good candidates for quantum memories, generation of such states can be of particular importance as it enables us to preserve the generated N00N states for the sufficiently long period of time. Therefore, in the following we are going to describe this architecture in details.

\subsection{N00N state in nitrogen vacancy centers in diamond}
Now, following Ref. \cite{maleki2018generating} we are going to show that N00N states can be generated in NVEs in diamond. The scheme of the N00N state generation is following the same N00N state generation circuit. Thus, we are going to sketch a scenario where a similar effective Hamiltonian 
can be extracted for NVEs interacting with a qubit.

\begin{figure}
\centering
\includegraphics[width=14 cm]{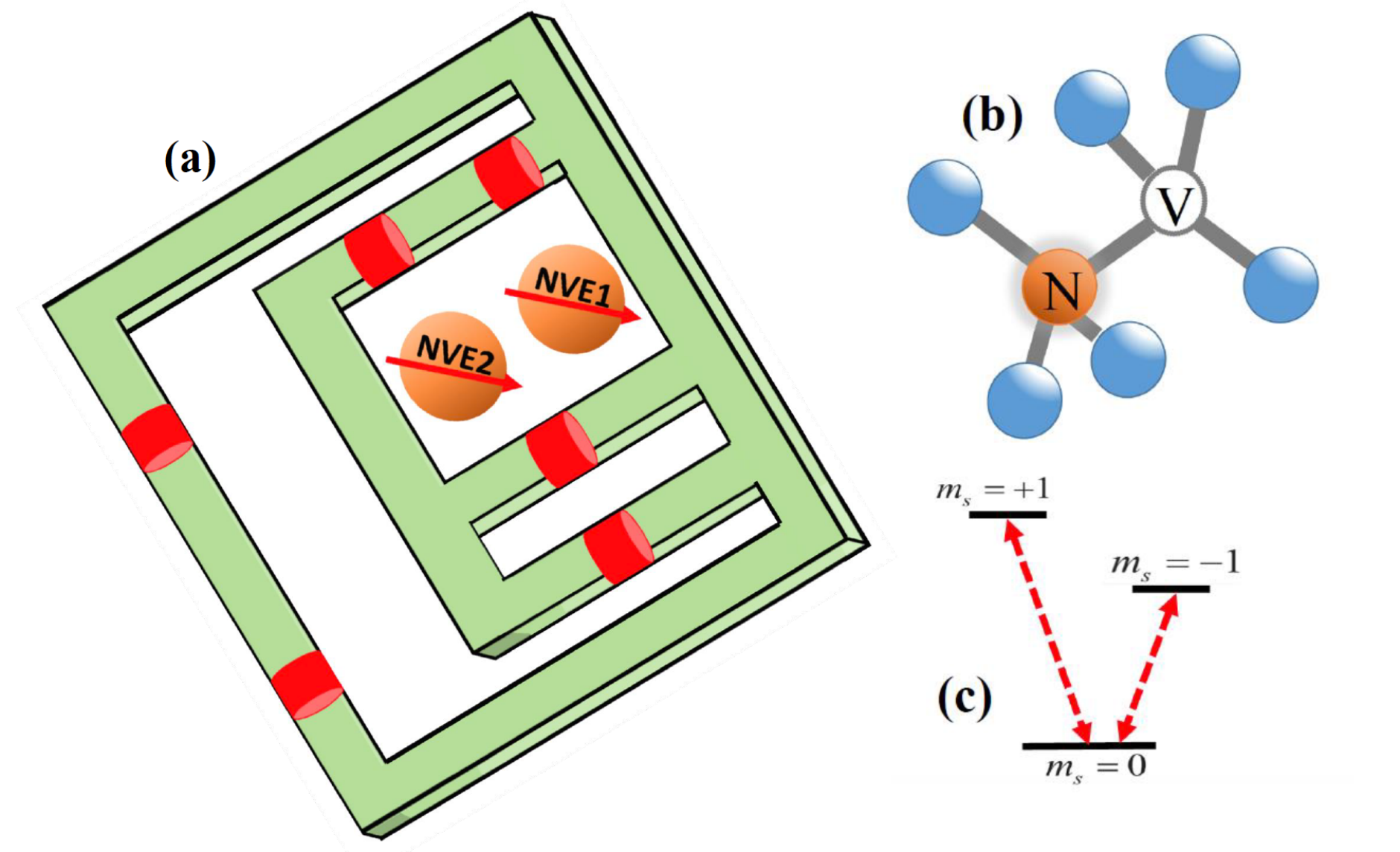}
\caption{
(a) Two ensembles of NV centers coupled to a superconducting flux qubit consisting of four Josephson junctions forming the main loop and an $\alpha$ loop. (b,c) Crystal-lattice (b) and energy (c) diagrams of an NV center in diamond.
}
\label{fig:Fig-M-3}
\end{figure}

In this platform, we  investigate a quantum architecture consisting of two separate noninteracting NVEs in diamond coupled to a common superconducting qubit. As a part of this architecture, we consider a large and gap-tunable flux qubit [Fig.\ref{fig:Fig-M-3}(a)] \cite{Marcos,Song}. 
Such a hybrid scheme combines the long coherence time of spins in NV centers, the advantage of controlling NV centers by microwave and optical fields \cite{Childress,Dutt}, the tunability property of superconducting devices, the circuit scalability, and the remarkable compatibility with the cutting-edge nanotechnologies \cite{Marcos,Song, 
Liu,Ranjan,Wang2}. It is notable that, even though microwave photons in superconducting resonators provide a unique platform for N00N state generation which was mentioned earlier, unlike microwave photons in superconducting resonators, which have lifetimes of the order of 1 ms \cite{Reagor}, the coherence time of  NV centers in diamond can approach 1 s even at relatively high temperatures \cite{Bar-Gill}. Thus, combining the advantages from both devices seems to be more desirable for the quantum state preparation purposes.

Now, we describe the quantum characteristics of the flux qubit. To this aim we note that we need to choose the biasing in the main loop of the flux qubit close to half the flux quantum, $\Phi_0=h/(2e)$. Therefore, the flux qubit could be described \cite{Song} in terms of a Hamiltonian in a two-dimensional Hilbert space. The Hamiltonian describing this system could explicitly be given by
$$
H=-[\epsilon(\Phi_{ext})\sigma_z+\Delta(\Phi_{ext}')\sigma_x]/2
$$
 where $\sigma_x $ and $\sigma_z$ are the well-known Pauli operators in the basis of flux qubit.  The parameter $\Delta(\Phi_{ext}')$ is the flux qubit tunneling splitting energy, and finally $\epsilon(\Phi_{ext})$ is the energy bias of the flux qubit. These parameters provide a platform for controlling the qubit Hamiltonian of the system.

When  a static magnetic field of half the flux quantum  is applied perpendicular to the main loop of the flux qubit, the clockwise and counterclockwise persistent current states in the system become almost degenerate \cite{Xiang2,Xiang3}. These clockwise and counterclockwise states provide us with the capability to define the logical basis states $|0\rangle_f$ and $|1\rangle_f$. In this setting,  $|0\rangle_f$ and $|1\rangle_f$ represent the states of the clockwise and counterclockwise persistent current, respectively. Noting the fact that $\epsilon(\Phi_{ext})$ and $\Delta(\Phi_{ext}')$ can be adjusted independently by external magnetic fluxes through the main and the $\alpha$ loops, one can set $\epsilon(\Phi_{ext})=0$. Therefore, by preparing the system in one of its computational basis $|0\rangle_f$ or $|1\rangle_f$, one can perform a single qubit operation on the system considering the Hamiltonian given above. In the following, we will show that such a flux qubit can be combined with NVEs to construct an interesting hybrid architecture for information processing and entangled state generation. To this aim we first need to consider the NVEs Hamiltonian in details.

 \par
 For the NV centers in diamond architecture, we only are interested in the ground state of such a system. Interestingly,
 the ground state of an NV center in diamond is a spin triplet state with zero-magnetic-field splitting of about $D_{gs}\approx$ 2.87 GHz between the $m_s=0$ sublevel and the degenerate $m_s=\pm1$ sublevels [Fig.\ref{fig:Fig-M-3}(b,c)]. To eliminate the degeneracy in the $m_s=\pm1$ sublevels, we can apply an external magnetic field to induce  Zeeman splitting. Therefore. we apply
 the external magnetic field $B_{ext}$ along the [100] direction of the diamond crystal lattice, to remove the degeneracy of $|m_s=\pm1\rangle$  sublevels [Fig.\ref{fig:Fig-M-3}(c)]. Hence, one can focus on the two sublevels $m_s=0$ and $m_s=-1$, to reduce the system into a two-level system. This enables us to reduce the Hilbert space of the system into a two-dimensional space in which one can define the computational basis for the information processing operations.
 With this description, the spin operators of the $j$th NV center can be defined though Pauli operators. Hence, we explicitly define
$$
s_{zj}=|-1_j\rangle\langle -1_j|-|0_j\rangle\langle 0_j|,
$$
$$
s_{+j}=|-1_j\rangle\langle 0_j|,
$$
$$
s_{-}=|0_j\rangle\langle -1_j|.
$$
The definition of the Pauli matrices for a single NV, enables one to define a generic scenario for the situation which includes  $N_0$ nitrogen vacancy centers. More precisely, if an NVE contains $N_0$ NV spin, collective spin operators can easily be expressed through summing the spins of the system which can be written as
 $$
 S_{\tau}=\sum_{k=1}^{N_0} s_{\tau k} (\tau=z,\pm).
 $$ 
 It is quite notable that qubits based on nitrogen vacancy centers with an external magnetic field $B_{ext}$ applied to induce splitting between the $m=1$ and $m=-1$ sublevels have been investigated  extensively in the literature (see, e.g., \cite{Maleki2018OE,Maleki2019JOSA, Xiang2, Xiang3}). One main difference of what we are going to present here with the earlier works is that we apply
  additional time-dependent $B_j$ fields,  to NVEs beside with $B_{ext}$ to provide a periodic modulation of the $m=0$ -- $m=-1$ splitting. This interestingly provides us with the capability of generating N00N states which is the main goal of our discussion in this chapter.
  To obtain the desired effective Hamiltonian, we are going to apply the Holstein--Primakoff (HP) transformation \cite{Hammerer} to the spin operators of NV centers. Such a strategy enables mapping of the collective spin behavior of a quantum system to a harmonics oscillator. Therefore, by applying HP transformation on the time-dependent $m=0$ and $m=-1$ splitting we show that the transformation leads to a closed-form Hamiltonian that we are seeking for.

To be able to apply the HP transformation, we need to restrict ourselves in the regime of weak excitation, where the spin operators of an NVE with a large $N_0$ can be mapped onto bosonic operators via an HP transformation \cite{Hammerer}. Thus, we assume the the the number of the excitation is very small compared to  $N_0$. With such an assumption, one can perform the HP transformation resulting in

$$
\sum_{k=1}^{N_0} s_{+k}^j=c_j^{\dag}\sqrt{N_0-c_j^{\dag}c_j}\backsimeq\sqrt{N_0}c_j^{\dag},
$$
and
$$
 \sum_{k=1}^{N_0} s_{-k}^j=c_j\sqrt{N_0-c_j^{\dag}c_j}\backsimeq\sqrt{N_0}c_j,
 $$
 also
  $$
  \sum_{k=1}^{N_0} s_{zk}^j=2c_j^{\dag}c_j-N_0.
  $$
  In the relations above $j=1,2$ for the first and the second NVEs, respectively. In this manner,one has $[c_j,c_j^{\dag}]=1$.
    
  We note that once we have a collective spin system, the coupling of the system enhances drastically compared to the single spin coupling. More precisely, the collective coupling strength of an ensemble of $N_0$ spins  enhances by a factor of $\sqrt{N_0}$ compared to coupling strength of a single spin \cite{Raizen}. Therefore, increasing  $N_0$ seems to to be favorable from this perspective as it is much easier to fulfill the requirements of the strong coupling regime.
  Furthermore, the weak-excitation requirement mentioned above is necessary for the validity of HP mapping, could be satisfied  via the conditions  $c_1^\dag c_1 <<N_0$ and $c_2^\dag c_2 << N_0$. This in turn limits the $N$ number in attainable N00N states, $N<<N_0$. However, $N_0$ can be sufficiently large in a realistic scenario. For instance, $N_0\sim$ $10^7$ has been used in the practicle experiments \cite{Zhu}, which is sufficiently higher than what we expect for the N00N state generations.

The total Hamiltonian of the NVE-flux-qubit hybrid system considered here can now be written as
 \begin{equation}\label{hamilton1}
H= -\hbar\frac{\Delta(\Phi_{ext}')}{2}\sigma_x+\hbar\sum_{j=1}^2\omega_j c_j^{\dag}c_j+\hbar\sum_{j=1}^2g(c_j^{\dag}+c_j)\sigma_z,
\end{equation}
 where 
 $$\omega_j=D_{gs}-g_e\mu_BB_z-g_e\mu_BB_j,
 $$
 
  $g_e$ is the ground-state Lande factor and $\mu_B$ is the Bohr magneton and $B_z$ is the magnetic field sensed by the spins due to the applied external magnetic filed $B_{ext}$ and the magnetic field produced by the flux qubit. Also, the parameter g is the coupling constant between the NVE esnenbles and the flux qubit.

 We Further assume that the sizes of the NVEs are sufficiently small to neglect spatial variations of the magnetic field induced by the flux qubit. Furthermore, the magnetic fields are chosen such that 
 $$
 \omega_j=\Delta(\Phi_{ext}')/2 + \delta \sin(\nu t+\varphi_j),
 $$
  where $\delta \sin(\nu t+\varphi_j)$ is controlled by the ac magnetic field. Choosing large $\nu$ and small $\delta$, so that $\Delta(\Phi_{ext}')\gg \zeta=\delta/\nu$, we apply the rotating-wave approximation (RWA) and use the basis of flux qubit eigenstates to reduce the interaction Hamiltonian of the considered system to 
\begin{equation}\label{}
H_I=\hbar g\sigma^+(\hat{c}_1 e^{i\zeta\cos(\nu t+\varphi_1)}+\hat{c}_2 e^{i\zeta\cos(\nu t+\varphi_2)})+H.c.
\end{equation}

This Hamiltonian can be reduced to a Floquet Hamiltonian.
We set $\zeta=2.40$ ($J_{0}(2.40)=0$) and $\varphi_2\neq\varphi_1$ to find the effective interaction Hamiltonian: 
\begin{equation}\label{hamilton2}
H\textsubscript{eff}= i\hbar \Omega(\hat{c}_1^\dag \hat{c}_2 - \hat{c}_2^\dag \hat{c}_1 )\sigma_z.
\end{equation}

It is straightforward to see now that the coupling coefficient can be tuned by varying the $B_j$ fields. In particular, with $\varphi_1-\varphi_2=\pi/3$ and $\nu=5 \chi g\simeq 3.14 g$, we have $\Omega/2\pi \approx 14$ MHz. Such coupling strengths have been recently demonstrated for a system of a flux qubit and an NVE with $N_0\approx$ $3\times10^7$ \cite{Zhu}. With the decay rates of NVEs and the flux qubit estimated as $\gamma_{NV}\sim$ 1 Hz \cite{Bar-Gill} and $\gamma_{FQ}\sim$ 1 MHz \cite{Lu}, we find that the requirement of strong coupling, $\Omega\gg   \gamma_{FQ}, \gamma_{NV}$, is fulfilled.

As one of its interesting features, our protocol of N00N-state generation requires only local operations on the flux qubit at the second and third steps, where the state $|\psi\rangle_1$ is degenerated into $|\psi\rangle_3$. Consequently, the operation time for each of these two steps reduces to the single-qubit operation time, $\Delta t_2$, $\Delta t_3 \sim$ 1 ns \cite{Xiang3}. This fact allows the overall time $\Delta \tau = \sum_{i=1}^4 \Delta t_i$ of N00N-state generation to be sufficiently shortened. This is useful for avoiding decoherence buildup in the system.  With our earlier estimate of $\Omega/2\pi\simeq14$ MHz, the system could be in a strong-coupling regime, for which we have $\Delta t_1\approx$ 9 ns. Hence, we find for the total time required to create a N00N state through the considered process to be $\Delta \tau \approx$ 20 ns when the $|0\rangle_f$ state is found at the measurement step and $\Delta \tau \approx$ 38 ns if the measurement yields $|1\rangle_f$. In both situations, the time that is needed for N00N-state generation is much shorter than the dephasing times of both the flux qubit, $\sim 1$ $\mu$s \cite{Lu} and the NVEs $\sim350$ $\mu$s \cite{Yang1}.

\subsection{N00N state as a superposition of light and motion}

We examine a trapped ion featuring an atomic transition between two electronic states, denoted as $|g\rangle$ and $|e\rangle$, with a transition frequency of $\omega_0$. This ion is coupled to a cavity field with a frequency of $\omega_{c}$ and is subjected to external stimulation through a classical laser field operating at the frequency $\omega_L$. In a rotating frame with the laser frequency $\omega_L$, the Hamiltonian is expressed as follows \cite{Parkins1,Parkins2} 
$$
H= \nu b^\dag b + \Delta_{c}a^\dag a+
\delta\sigma_z+\varepsilon_L\sigma_++\varepsilon^*_L\sigma_-
$$
$$
+g_0\
\sin\eta(b^\dag +b)(a^\dag \sigma_-+a\sigma_+).
$$

In this context, the symbols $a$, $a^\dag$, $b$, and $b^\dag$ represent the bosonic annihilation and creation operators for the cavity field and the quantized atomic vibration, respectively. The operators $\sigma_z= |e\rangle\langle e| - |g\rangle\langle g|$, $\sigma_+= |e\rangle\langle g|$, and $\sigma_-= |g\rangle\langle e|$ are the Pauli operators. The Lamb-Dicke parameter corresponding to these operators is denoted by $\eta$. The detuning parameters, namely $\Delta_{c}$ and $\delta$, are defined as $\Delta_{c}=\omega_c-\omega_L$ and $\delta=\omega_0-\omega_L$.

The parameter $\varepsilon_L = \epsilon_L e^{-i\varphi_L}$ represents the amplitude of the laser field. The coupling strength between a single photon, the atom, and the cavity is denoted by $g_0$, and the sinusoidal function captures the standing wave pattern of the cavity field.

 In the Lamb-Dicke regime $(\eta \ll 1)$, we can simplify $\textrm{sin }\eta(b^\dag + b) \approx \eta(b^\dag + b)$ \cite{Parkins1,Parkins2}. Under the resonant condition $\delta = 0$ and in the interaction picture with $\nu = \Delta_{c}$, after applying the rotating wave approximation (RWA), the reduced Hamiltonian takes the form
$$
H'= \varepsilon_L\sigma_++\varepsilon^*_L\sigma_-
+g_0\eta(b a^\dag \sigma_-+b^\dag a\sigma_+).
$$
We transition to the transformed picture $H_I=RH'R^\dag$ through the unitary transformation
$$
R=e^{i \frac{\pi}{4}\sigma_y}e^{-i\frac{\varphi_ L}{2} \sigma_z}e^{i \epsilon_l \sigma_zt}
$$
Where, following the adiabatic elimination of rapidly oscillating terms using the Rotating Wave Approximation (RWA) with the condition $g_0 \eta \ll \epsilon_L$, and introducing $\varphi_L\rightarrow\varphi-\frac{\pi}{2}$, the resulting effective interaction Hamiltonian is given by \cite{Maleki2018JOSA}

\begin{equation}\label{hamilton}
H_{eff}= i\Omega(e^{-i\varphi}a^\dag b - e^{i\varphi}a b^\dag)\sigma_z,
\end{equation}
with $\Omega=g_0 \eta/2$.

\section{Entangled state generations beyond the N00N state}

Here, we are going to consider generation of entangled state beyond the N00N state. These entangled states can be very useful for various quantum information processing tasks \cite{Maleki2019LPL, Maleki2019LOL}. We are going to show that a vast class of entangled states can be generated with the scheme that will be provided here. More interestingly, various maximally entangled states can be engineered with the protocol. We are going to follow \cite{maleki2019high} in doing so. The starting point to our system is still the same effective  Hamiltonian that we have already focused on. In this platform, we consider two cavities which interact with a common qubit. Using the Hamiltonian in Eq. (\ref{N00Nham1})
we can derive the dynamics of the field operators as
\begin{align*}
\dot{a}_1&=\kappa a_2(t)\sigma_z,
\\
\dot{a}_2&=-\kappa a_1(t)\sigma_z.
\end{align*}
The solution to these equations can be found to be
\begin{align*}
a_1= a_1(0) \cos(\kappa t)-a_2(0) \sin(\kappa t)\sigma_z,
\\
a_2= a_2(0) \cos(\kappa t)+a_1(0) \sin(\kappa t) \sigma_z.
\end{align*}
At $\kappa t={\pi}/{4}$, we have
\begin{align*}
a_1= \frac{1}{\sqrt{2}}(a_1(0) -a_2(0) \sigma_z),
\\
a_2=\frac{1}{\sqrt{2}}( a_2(0) +a_1(0)\sigma_z).
\end{align*}
Next, we let the photons to go through a 50/50 beam splitter given by the two--mode unitary operation 
$$
U=\text{exp}[-\pi/4( a_1^{\dag}(0)a_2(0)- a_2^{\dag}(0)a_1(0))].
$$
 This process results in the following transformation of the field operators
\begin{align*}
a_1(0)= \frac{1}{\sqrt{2}}(a_1(0) +a_2(0) ),
\\
a_2(0) =\frac{1}{\sqrt{2}}( a_2(0) -a_1(0)).
\end{align*}
Thus, the total transformation, after passing through the beam splitter can be given by 
\begin{align}
a_1 \rightarrow a_1(0) |e\rangle\langle e| +a_2(0) |g\rangle\langle g|,
\\ 
a_2 \rightarrow a_2(0) |e\rangle\langle e| -a_1(0) |g\rangle\langle g|
\end{align}

These transformations are central to generation of entangled states which will become clear soon. Note that these are conditional photon transformation subjected to the state of the qubit. We will see that such a transformation maps various initial states into interesting entangled states which could be useful for various tasks. 
Now, to see the power of such a transformation, let us consider an specific example as the input state 
$$
|\psi\rangle=|N\rangle
|0\rangle \frac{1}{\sqrt{2}}( |g\rangle +|e\rangle).
$$
 The above transformation, acting upon this initial state, gives rise into
$$
|\psi\rangle \longrightarrow\frac{1}{\sqrt{2}}(|N\rangle |0\rangle  |e\rangle+ |0\rangle |N\rangle  |g\rangle).
$$
\begin{figure}
\includegraphics[width=\columnwidth]{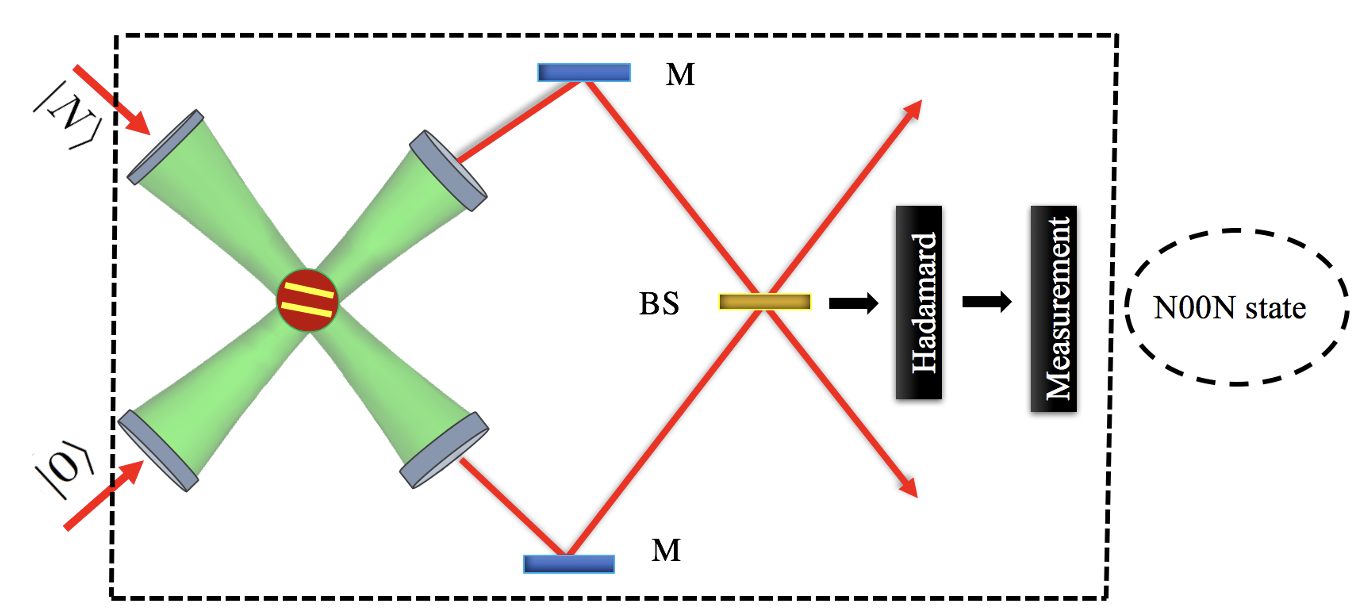}
\caption{Architecture of the entangled state generation setup. The input state first interacts with the system of the two resonators interacting with a qubit for as long as  $\kappa t={\pi}/{4}$. Then the photons in the resonators are sent to a beam splitter. Finally, by applying a Hadamard gate and performing a measurement on the qubit we attain a N00N state.
}
\label{fig:Fig-M-5}
\end{figure}
By applying an external $\pi/2$ pulse (Hadamard gate) to the atom in the state above the overall state of the system degenerates into
$$
[(|N\rangle
|0\rangle +|0\rangle |N\rangle) |e\rangle+(|N\rangle
|0\rangle -|0\rangle |N\rangle) |g\rangle]/2.
$$
Therefore, the measurement on the state of the atom provides different N00N states. Namely, detection of the atom in the exited state results in the N00N state 
$$
\frac{1}{\sqrt{2}} (|N\rangle
|0\rangle +|0\rangle |N\rangle),
$$
 and detection of the atom in its ground state gives 
 $$\frac{1}{\sqrt{2}} (|N\rangle
|0\rangle -|0\rangle |N\rangle).
$$ 
Regardless of the sign differences, both of the states give the same result in the phase estimation scenario, and both are N00N states.
We summarize the transformation presented here in Fig. \ref{fig:Fig-M-5}. Accordingly, the input state first interacts with the resonators--qubit system, then goes through a beam splitter. Finally, a Hadamard gate followed by a measurement is performed on the qubit which gives the desired N00N states.

As was mentioned earlier, the usefulness  of our architecture can even be extended beyond the N00N state generation. For instance, let us consider the case where we  prepare the initial state of the first cavity in a coherent state \cite{Scully,Maleki2017}
$$
|\alpha\rangle=e^{-\frac{|\alpha|^2}{2}}\sum_{n=0}^\infty
\frac{\alpha^n}{\sqrt{n!}}|n\rangle,
$$
where $\alpha$ is an arbitrary complex number. 
To this aim, we consider the initial state of the system to be expressed as  
$$
|\psi\rangle=|\alpha\rangle
|0\rangle \frac{1}{\sqrt{2}}( |g\rangle +|e\rangle),
$$
 where  we could generate the superposed state of the form
$$
|\psi\rangle \longrightarrow\frac{1}{\sqrt{2}}(|\alpha\rangle |0\rangle  |e\rangle+ |0\rangle |\alpha\rangle  |g\rangle)
$$
This is a micro-macro entangled state where an atom is entangled with a shared coherent state in the cavities. Thus, one can generate entangled state of the form 
$$
|\varphi^\pm\rangle=\frac{1}{\sqrt{2\pm2 e^{-|\alpha|^2}}} (|\alpha\rangle
|0\rangle \pm|0\rangle |\alpha\rangle).
$$
The state $|\varphi^+\rangle$ is proven to be very usefull in quantum merology \cite{Joo}.  If we start with  a squeezed state at the input stage, we could entangle the squeezed state with the vacuum.

Our approach can generate states even beyond such a superposition. In general, we take the quantum states 
$$
|\psi_1\rangle=\sum_{n=0}^{d_1}b_n|n\rangle,
$$ 
and 
$$
|\psi_2\rangle=\sum_{n=0}^{d_2}c_n|n\rangle.
$$
 Hence, we prepare the initial state 
in  
$$
|\psi\rangle=|\psi_1\rangle
|\psi_2\rangle \frac{1}{\sqrt{2}}( |g\rangle +|e\rangle).
$$ 
 Then, given the detection of the atom in its exited state, we can generate the symmetric  entangled state
\begin{equation}
|\psi^+\rangle = \dfrac{1}{\sqrt{2+ 2|\langle \psi_1 |\psi_2\rangle|^2}}(|\psi_1\rangle |\psi_2\rangle + |\psi_2\rangle |\psi_1\rangle),
\end{equation}
alternatively, if we detect the atom in its ground state, we attain the anti-symmetric entangled state
\begin{equation}
|\psi^-\rangle = \dfrac{1}{\sqrt{2- 2|\langle \psi_1 |\psi_2\rangle|^2}}(|\psi_1\rangle |\psi_2\rangle - |\psi_2\rangle |\psi_1\rangle).
\end{equation}
Thus, we can generate several interesting states in our architecture such as superposition of  two coherent states $|\alpha\rangle$ and $ |\beta\rangle$
$$
\dfrac{1}{\sqrt{2\pm 2|\langle \alpha |\beta\rangle|^2}} (|\alpha\rangle |\beta\rangle \pm |\beta\rangle |\alpha\rangle),
$$
or superposition of  two squeezed states $|\xi_1\rangle$ and $ |\xi_2\rangle$
$$
\dfrac{1}{\sqrt{2\pm 2|\langle \xi_1 |\xi_2\rangle|^2}} (|\xi_1\rangle |\xi_2\rangle \pm |\xi_2\rangle |\xi_1\rangle),
$$
or even hybrid entangled states of the forms
$$
\dfrac{1}{\sqrt{2\pm 2|\langle \xi_1 |\alpha\rangle|^2}}(|\alpha\rangle |\xi_1\rangle \pm |\xi_1\rangle |\alpha\rangle),
$$
$$
\dfrac{1}{\sqrt{2\pm2|\langle N |\alpha\rangle|^2}}(|\alpha\rangle |N\rangle \pm |N\rangle |\alpha\rangle),
$$
$$
\dfrac{1}{\sqrt{2\pm 2|\langle N |\xi_1\rangle|^2}}( |\xi_1\rangle |N\rangle \pm |N\rangle  |\xi_1\rangle).
$$

Such states are of paramount importance in beating the shot noise limit in quantum metrology and  have a vast application in  quantum information processing.

To quantify the entanglement degree of the generated states $|\psi^\pm\rangle$
we can use concurrence as the measure of entanglement \cite{Wootters},
which is defined as
\begin{equation}
C=|\langle\psi|\sigma_{y}\otimes\sigma_{y}|\psi^*\rangle|,
\end{equation}
where $\sigma_{y}$ is the spin flip operator and $|\psi^*\rangle$ is
the complex conjugate of $|\psi\rangle$.
With some straightforward calculation, we find the  concurrence as
\begin{equation}
C=\dfrac{1-|\langle \psi_1 |\psi_2\rangle|^2}{1\pm |\langle \psi_1 |\psi_2\rangle|^2}.
\end{equation}
Where, the plus(minus) sign in the denominator shows the concurrence of the state $|\psi^+\rangle$($|\psi^-\rangle$).
Interestingly, $|\psi^-\rangle$ is maximally entangled for any input states. However, $|\psi^+\rangle$ is maximally entangled if and only if the input states in the modes are orthogonal states, i.e., $\langle \psi_1 |\psi_2\rangle=0$.
For instance, considering the input state $|N\rangle
|M\rangle$, one can generate maximally entangled states

$$
|\psi\rangle \longrightarrow\frac{1}{\sqrt{2}}(|N\rangle |M\rangle+ |M\rangle |N\rangle),
$$
in this class of states. This reduces to the N00N states when $M=0$.

Considering  the maximal entanglement of $|\psi^-\rangle$, a wast class of maximally entangled states can be generated in this setup. It is notable that, having maximally entangled states are extremely important in many quantum protocols. This scheme can be useful in a vast span of quantum disciplines and can provide a new platform for information processing tasks.

\section{Eantangled state generation beyond the two-mode states}

As was shown earlier, it is quite possible to generate a large class of entangled states, such as N00N state and entangled coherent states in the architecture that we have described earlier. It would be quite interesting to extend the scheme to a miltimode scenario. Generation of entangled states with more than two modes could be very important for various applications as well. However, generation of multimode quantum states is one of the most challenging tasks in most of the  practical applications of quantum science. Thus, what will be presented here could be found useful in various quantum disciplines.

\begin{figure}
\center
\includegraphics[width=13 cm]{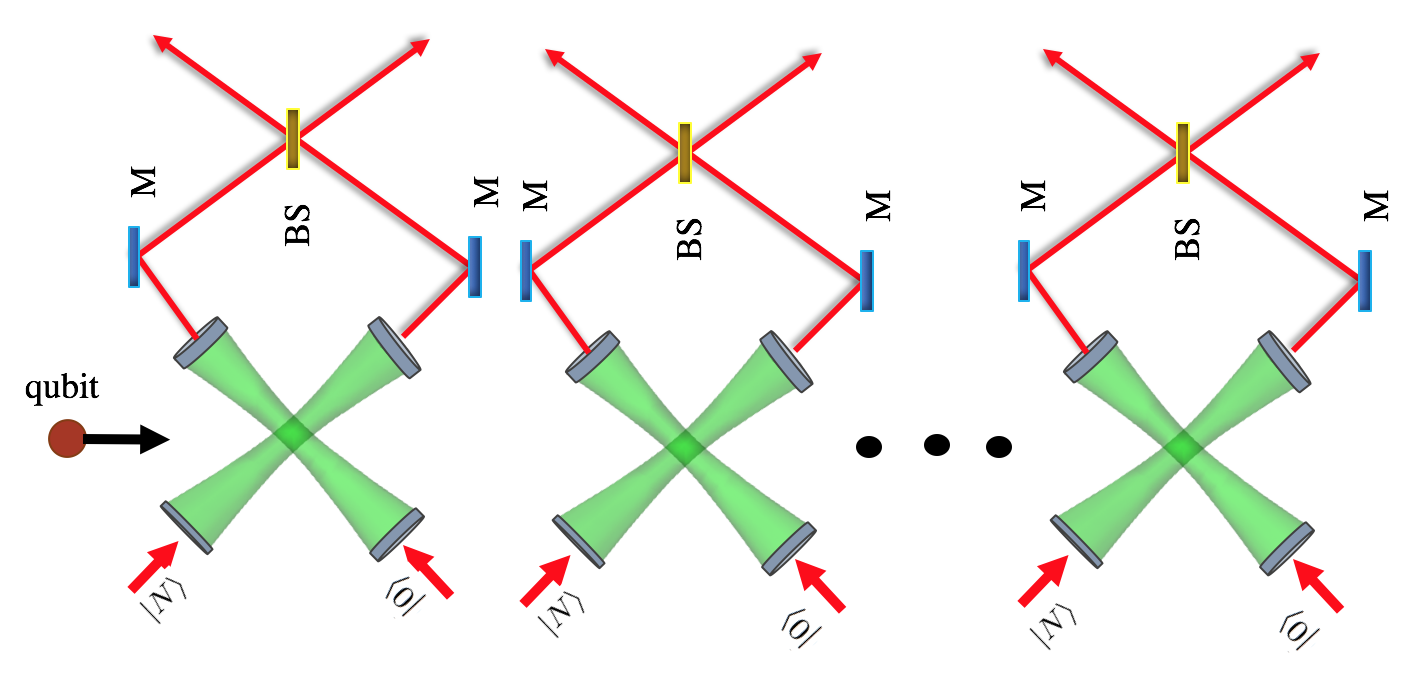}
\caption{The network of multi-N00N state generation architecture. The qubit is sent through the resonator network to interact with each resonator pair. Then the photons inside the each pair is sent to a beam splitter. 
}
\label{fig:Fig-M-6}
\end{figure}

Now, we would like to show how the scheme that we investigated earlier, can be used to generate a multimode state. To do so. we first prepare an array of two-cavity system where, the array consists of pairs of cavities and a single qubit(see Fig. \ref{fig:Fig-M-6}). Furthermore, we  prepare the initial state of the entire system in some specific state of the interest. For instance, let us examine the following initial state as an input of the scheme
$$
|\psi\rangle=|N,0\rangle\otimes
|N,0\rangle\otimes ... \otimes|N,0\rangle \frac{1}{\sqrt{2}}( |g\rangle +|e\rangle).
$$
 Then, we send the qubit through the two-resonator network, where it interacts for the time length of  $\kappa t={\pi}/{4}$ with each resonator pair.  Once the qubit is interacted with each pair, the photons go though the beam splitters after that. Consequently, the initial state of the system degenerates into
 $$
|\psi\rangle \longrightarrow\frac{1}{\sqrt{2}}(|N,0\rangle ...|N,0\rangle  |e\rangle+ |0,N\rangle ...|0,N\rangle  |g\rangle).
$$
Now, by applying a ${\pi}/{2}$ pulse on the qubit and performing a measurement we get the multi-N00N state 
 $$
|\psi\rangle \longrightarrow\frac{1}{\sqrt{2}}(|N,0\rangle ...|N,0\rangle  \pm|0,N\rangle ...|0,N\rangle ).
$$

It is interesting that for an array of $M$ resonator pairs, the phase resolution of the system reduces to $\Delta\varphi\propto\frac{1}{M N}$. To achieve this value through a N00N state we should be able to increase the number of phtons to $M N$ in each of the entangled modes which is a very hard task for relatively large $M$ and  $N$.

Recently, a scheme to generate double N00N states  was proposed \cite{Su2}. To appreciate, the approach we introduce here, we restrict ourselves to $M=2$ and compare the proposal of our work with the approach of \cite{Su2}. In their setup they need $N+2$ operational steps to generate a double N00N state; however, in our system we can generate a N00N state in just a few operational steps, which is independent of the number of the photons. This is a big advantage of our approach. Furthermore, our structure is even capable of generating multi-N00N states that can facilitate the quantum phase estimation techniques. That being said, we can generate a vast class of multi-mode entangled state by preparing each resonator in our specific state of interest.

\section{Heisenberg-Limit metrology beyond N00N states}

It was already discussed that N00N states have the specific capabilities to reach the Heisenberg-Limit for the phase estimation sensitivity. Here, we discuss the HL sensitivity in the su(2) algebra representation and discuss some of its properties. In fact, what we are going to show is that there is a large class of Fock state superposition which can reach the HL sensitivity. To lay out a rather general formalism for such states we will consider the relation between N00N state and spin coherent state superposition based on \cite{maleki2020spin}. To this aim, we note that for such a two-mode Fock state representation, one may introduce the Schwinger realization of the su(2) algebra in terms of the field operators as \cite{Maleki2015}
$$
J_+=a^\dag b,\qquad J_- =b^\dag a,\qquad J_z =\frac{1}{2}(a^\dag a -
b^\dag b).
$$
Hence, $J_-$ and $J_+$ are the lowering and raising operators of the
spin states, respectively. The generators of su(2) Lie Algebra
$J_\pm$ and $J_z$ satisfy the commutation relations 
$$
[J_+,J_-]=2J_z,\qquad [J_z,J_\pm]=\pm J_\pm.
$$
Taking $j=\frac{n}{2}$, and $m=\frac{n_a-n_b}{2}$, one can define
$$
|n_a\rangle_a |n_b\rangle_b\equiv |j,m\rangle, \qquad (n=n_a+n_b).
$$
This relation maps all the basis of the Hilbert space of the entire system from Fock state products to the Dicke state basis described by the su(2) algebra. The two special cases which are interesting to note are given by  
$$
|0\rangle_a \otimes|2j\rangle_b\equiv
|j,-j\rangle, \quad \text{and } \quad |2j\rangle_a \otimes|0\rangle_b\equiv
|j,j\rangle.
$$ 

 With this definition, the generators of the spin operators act on the basis as 
$$
J_\pm|j,m\rangle=\sqrt{(j\pm m)(j\pm m+1)} |j,m\pm1\rangle,\quad \text{and} \quad J_z|j,m\rangle=m|j,m\rangle.
$$ 

Therefore, given $N$ photons in  one of the two modes, the two special cases could be expressed as 
$$
|0\rangle_a \otimes|N\rangle_b\equiv
|j,-j\rangle_z, \quad \text{and} \quad |N\rangle_a \otimes|0\rangle_b\equiv
|j,j\rangle_z.
$$
These  realizations in hand,  one can rewrite the N00N state in Dicke basis as \cite{Sanders}
$$
|\textrm{N00N} \rangle= \frac{1}{\sqrt{2}} (|j,j\rangle_z +|j,-j\rangle_z).
$$
Thus, the N00N state is the superposition of the two north and the south poles of the Bloch sphere in the Dicke state basis. We note that, the superposition of the states with other principle numbers, i.e., 
$$
\frac{1}{\sqrt{2}} (|j,m\rangle_z +|j,-m\rangle_z),
$$ 
cannot give raise to the N00N state since in the Fock basis it is equivalence to the state 
$$
\frac{1}{\sqrt{2}} (|\frac{n}{2}+m\rangle_a|\frac{n}{2}-m\rangle_b +|\frac{n}{2}-m\rangle_a|\frac{n}{2}+m\rangle_b).
$$
Clearly, such state could not reproduce the HL sensitivity that is achievable with N00N state. However, considering the fact that N00N state are the two extreme points on the $z$ direction, one might think of considering the states in some other extreme points such as points in the $x$ direction. In fact, the superposition of the two extreme points in  $x$ direction could be expressed as 
$$
|j,\pm j\rangle_x=\frac{1}{2^j}\sum_{m=-j}^{j}
 \left(
  \begin{matrix}
    2j  \\
    j+m 
  \end{matrix}
  \right)^\frac{1}{2}(\pm1)^{j+m}|j,m\rangle_z,
$$
Noting that $|j,m\rangle_z=|\frac{n}{2}+m\rangle_a|\frac{n}{2}-m\rangle_b$, the superposition of such states is not equal to the N00N state. In fact, the normalized superposition of the two  extreme points in  $x$ direction is
$$
|\psi \rangle= \frac{1}{\sqrt{2}} (|j,j\rangle_x +|j,-j\rangle_x).
$$
Now, we show that this state could be equivalent to the N00N state in phase estimation scenario if we perform an specific strategy that will become clear soon.
To see this, we need to show that this state can indeed give Heisenberg limit precision. For this purpose, let us apply the unitary operator around $x$-axis $\exp(i\theta_x j_x)$ on the state $|\psi \rangle$. With this operator applied, the state $|\psi \rangle$ becomes 
$$
|\psi (\theta_x)\rangle_= \frac{1}{\sqrt{2}} (e^{i\theta_x j}|j,j\rangle_x +e^{-i\theta_x j}|j,-j\rangle_x).
$$
Therefore, we can calculate the phase uncertainty as $\Delta\theta_x=\frac{1}{2j}=\frac{1}{N}$. Thus, the phase estimation  is exactly equal to the HL as the N00N state. From this result one can easily realize that the superposition of the two lowest and highest  states in the $y$ direction, i.e., $
|j,\pm j\rangle_y$ can reach the HL as well. These considerations through the extreme points in the $x$ and the $y$ directions was clearly pointed investigated in Ref. \cite{Sanders}. 

\begin{figure}
\centering
\includegraphics[width=12 cm]{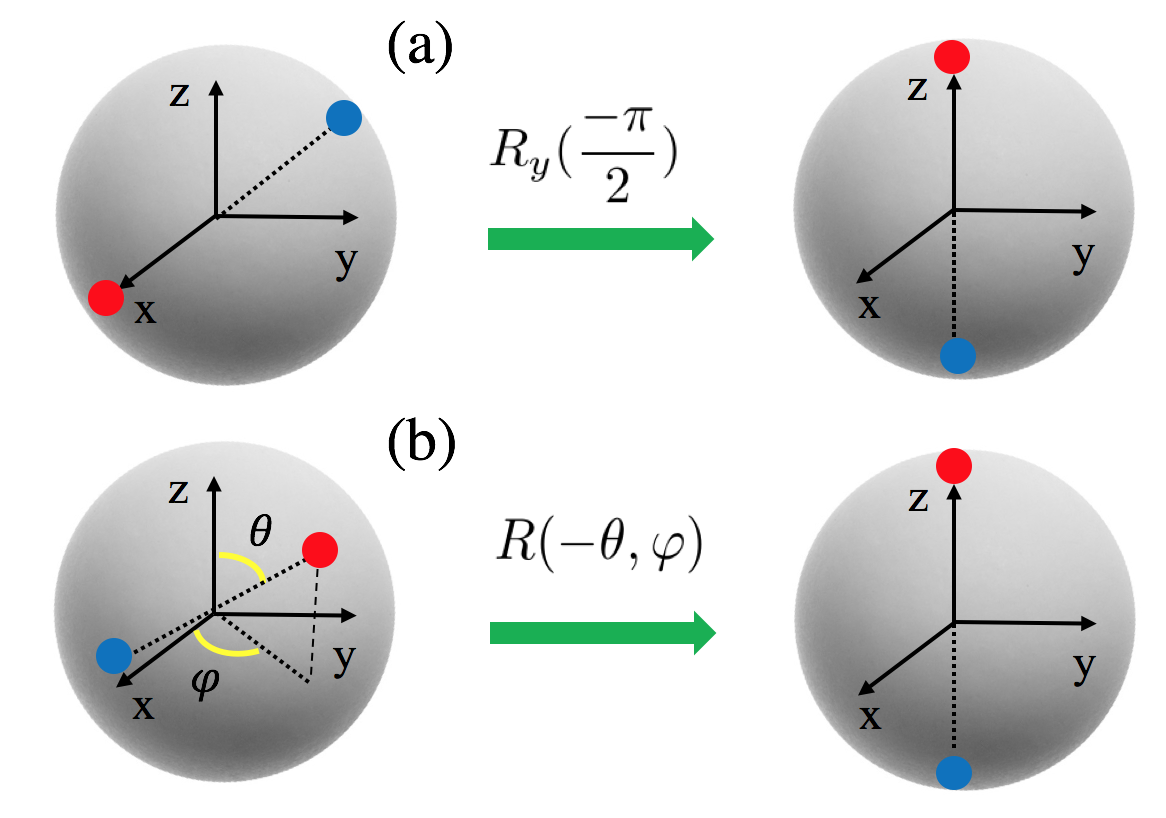}
\caption{
Rotation of the $|j,\pm j\rangle_x$ about y-axis on the Bloch sphere.
}
\label{Fig-M-7}
\end{figure}

Now, it is tempting to ask that if there is any other superposed state that can reach this limit, or are these three states the only ones that can fulfill the N00N state property for the precise measurement? If there are some other states what are these states? Also, how we can generate such superposition states of the Dicke basis say in $x$, $y$ and $z$ directions?  We shall address the generation scheme later; however, we first are going to discuss the first questions.  

It is remarkable that in such a cat state the two superposed points are antipodals on the Bloch sphere. In general, an element of the Bloch sphere can be obtained by applying a rotation operator on a reference point (such as $\vert j,-j\rangle_z$). Such rotation can be given based on the generators of the su(2) algebra of the spin system. The most general form of the rotation operator then can be expressed as \cite{Maleki2015}
$$
R(\theta,\varphi)  e^{i\vartheta J_z}=\exp{\lbrace\frac{-\theta}{2}(J_+
e^{-i\varphi}-J_- e^{i\varphi})\rbrace} e^{i\vartheta J_z}.
$$
On the $z$ direction basis, the term $e^{-i\vartheta J_z}$ give rise to an overall phase, that dose not effect the precision measurement. Since, The state
$$
\frac{1}{\sqrt{2}} (e^{i\vartheta_1}|N\rangle_a
\otimes|0\rangle_b +e^{-i\vartheta_1}|0\rangle_a \otimes|N\rangle_b),
$$
is as good as the N00N state, for precise measurement purposes (this extra phase degree of freedom was introduced in Ref.\cite{Sanders} to facilitate the generation of N00N state). However, the first term displaces the ground state of the Dicke basis to a coherent state as
\begin{align}\label{spincoh}
|\theta,\varphi, j\rangle=\exp{[\frac{\theta}{2}(J_+
e^{-i\varphi}-J_- e^{i\varphi})]} |j,-j\rangle
=\frac{1}{(1+|\gamma|^2)^j}\sum_{m=-j}^j
\left(
  \begin{matrix}
    2j  \\
    j+m 
  \end{matrix}
  \right)^\frac{1}{2}\gamma^{j+m}|j,m\rangle,
\end{align}
where $\gamma=e^{-i\varphi}\tan(\frac{\theta}{2})$. The overlap of two spin
coherent states is given by
$$
\langle \delta, j|\gamma, j\rangle=\frac{(1+\bar{\delta}
\gamma)^{2j}}{(1+|\delta|^2)^j(1+|\gamma|^2)^j}.
$$ 
Now, we examine the superposition of two antipodal coherent states on the Bloch sphere. We consider, with out loss of generality, a superposition given by \cite{Huang2015,Huang2018}
$$
 \frac{1}{\sqrt{2}} (|\theta,\varphi, j\rangle+|\pi-\theta,\pi+\varphi, j\rangle),
$$
In order for to investigate the phase shift and sensitivity obtained through this superposition, we consider the following Hermitian operator.
$$
J_\gamma=J_z \cos\theta-\frac{1}{2}(J_+
e^{-i\varphi}+J_- e^{i\varphi})\sin\theta.
$$
In fact, with such a Hermitian operator, we can define the unitary  phase shift operator $U=e^{-i\xi J_\gamma}$, such that
\begin{align}\label{}
e^{-i\xi J_\gamma} \frac{1}{\sqrt{2}} (|\theta,\varphi, j\rangle+|\pi-\theta,\pi+\varphi, j\rangle)
=\frac{1}{\sqrt{2}} (e^{i\xi j}|\theta,\varphi, j\rangle+e^{-i\xi j}|\pi-\theta,\pi+\varphi, j\rangle).
\end{align}
Thus, the phase uncertainty can be obtained as $\Delta\xi=\frac{1}{2j}$. Therefore, we can get HL phase uncertainty using the superposition of the two orthogonal spin coherent states as above. This shows that the HL phase uncertainty is not an exclusive property of the N00N states, and it can indeed be extracted with a large class of quantum states.

Now, one may ask whether it is possible to get such HL precision via superposition of two non-orthogonal spin coherent states which are not antipodal points or is this just limited to the antipodal orthogonal coherent state superposition? To have a better insight into such superposition we show the relation between the N00N state and such superposition states.
To this aim we note that the two extreme points of the $x$ direction are related to the diagonal $z$ component of the spin coherent state in Eq. \ref{spincoh} by the
$\frac{\pi}{2}$ rotation $R_y(\frac{\pi}{2})=\exp(-i\frac{\pi}{2}j_y)$ about the y-axis, i.e., 
$$
R_y(\frac{-\pi}{2})|j,\pm j\rangle_x =|j,\pm j\rangle_z.
$$
 This is shown in Fig. \ref{Fig-M-7}. 
Thus, by rotating the cat state about y-axis and noting that $|0\rangle_a|N\rangle_b\equiv
|j,-j\rangle_z$, and $|N\rangle_a|0\rangle_b\equiv
|j,j\rangle_z$ we create the N00N state
$$
|\textrm{N00N} \rangle= \frac{1}{\sqrt{2}} (|2j\rangle_a
\otimes|0\rangle_b +|0\rangle_a \otimes|2j\rangle_b).
$$

This result clearly shows that a N00N state, is in fact a
typical type of spin cat state \cite{Sanders}.
A coherent state can be expressed as an action of the rotation operator on the ground state such that $|\theta,\varphi, j\rangle=R(\theta,\varphi)|j,-j\rangle$. It is also remarkable that the antipodal point of this coherent state is 
$$
|\pi-\theta,\pi+\varphi, j\rangle= R(\theta,\varphi)R(\pi,\varphi)|j,-j\rangle= R(\theta,\varphi)|j,j\rangle.
$$
 Therefore, by applying the rotation operator $R^{-1}(\theta,\varphi)=R(-\theta,\varphi)$, we can get the N00N state. Thus, the spin cat states of antipodal points can be converted to the N00N state by some global unitary transformation. It is notable that such transformations change the states entirely and may transform it to a completely different state. With this description, now consider a superposition of two spin coherent state as $
|\theta_1,\varphi_1, j\rangle+|\theta_2,\varphi_2, j\rangle$. We apply the operator $R(-\theta_1,\varphi_1)$ on the state and noting that the product of two rotation operators gives a rotation operator, up to a phase factor, we may express the state as
$$
|j,-j\rangle+R(-\theta_1,\varphi_1)R(\theta_2,\varphi_2)R(\pi,0)|j,j\rangle.
$$

 This can be fulfilled for $R(\theta_2,\varphi_2)=e^{i\beta}R(\theta_1,\varphi_1)R(\pi,\varphi')$. Where we have used the fact that  $|j,-j\rangle=R(\pi,0)|j,j\rangle$. Now considering the fact that multiplication of two rotation operators gives rise to a rotation operator, up to a phase factor
$$
R(\gamma_1) R(\gamma_2)=R(\gamma_3) e^{i\Phi(\gamma_1,\gamma_2)J_z}
$$
where 
$$
\gamma_3=\frac{\gamma_1+\gamma_2}{1-\gamma_1^*\gamma_2},\qquad 
\Phi(\gamma_1,\gamma_2)=-i \ln ( \frac{1-\gamma_1^*\gamma_2}{1-\gamma_1\gamma_2^*} )
$$
with  $\gamma_1=e^{-i\varphi_1} \tan (\frac{\theta_1}{2})$, and  $\gamma_2=e^{-i\varphi_2} \tan (\frac{\theta_1}{2})$.
Thus, to have a N00N state one needs to have 
$$
R(-\theta_1,\varphi_1)R(\theta_2,\varphi_2)R(\pi,0)=I e^{i\Theta J_z}
$$
Where, $\Theta$ is some phase determined by the variables of $R(\theta_1,\varphi_1)$ and $R(\theta_2,\varphi_2)$. Thus, with this description the states
$|\theta_1,\varphi_1, j\rangle$, and $|\theta_2,\varphi_2, j\rangle$ need to be antipodals. In other cases the action of the operator $e^{i\phi J_z}$ shall not result in the HL metrology. We may express this in Fock basis as
\begin{equation}
\frac{1}{(1+|\gamma|^2)^\frac{n}{2}}\sum_{m=0}^n
\left(
  \begin{matrix}
    n  \\
    m 
  \end{matrix}
  \right)^\frac{1}{2}\lbrace\gamma^{m}+e^{i \psi}(-\frac{1}{\bar{\gamma}})^{m}\rbrace
 |m\rangle_a|n-m\rangle_b,
\end{equation}
where $n$ is the number of the total photons. It is quite evident that this state is completely different from the N00N state, however, it can give the HL precision as N00N state. Therefore, such a two-mode superposition of the photon number states can give the set of the states satisfying this limit. This may allow us to circumvent the difficulty of generating of the N00N state and permits more general ways to produce the state that may reach HL precision.

The interaction Hamiltonian of the two-mode field and the two-level atom reads
\begin{equation}\label{hamilton}
H_1= i\hbar\Omega(a^\dag b - b^\dag a)\sigma_z
\end{equation}

Now if the atomic state is initially prepared at $
\frac{1}{\sqrt{2}}(|g\rangle+|e\rangle)$ and one of the fields in
the Fock state $|n\rangle_b$, and the other is prepared in the
vacuum state $|0\rangle_a$, then 
$$
|\psi(0)\rangle= \frac{1}{\sqrt{2}}(|g\rangle+|e\rangle)|0\rangle_a
|n\rangle_b.
$$
Therefore,
defining $\xi(t)=e^{i\varphi}\tan(\Omega t)$ the initial state evolves to
$$
|\psi(t)\rangle= \frac{1}{\sqrt{2}} (|e\rangle\otimes |\xi(t),
j\rangle + |g\rangle\otimes |-\xi(t), j\rangle).
$$
This, in fact, is a type of micro-macro entanglement between the
electronic state of the atom and the cavity fields. We apply a
$\frac{\pi}{2}$ pulse on the atomic state which transforms
$|e\rangle\rightarrow \frac{1}{\sqrt{2}}(|g\rangle+|e\rangle)$ then
make a measurement on the ground state of the electronic state
 that projects the state to
$
 \frac{1}{\sqrt{2}} (|\xi(t), j\rangle + |-\xi(t),
j\rangle),
$
which is a spin cat state. Therefore, there is a large class of spin cat states that can give HL metrology.

\section{SUMMARY}

In this review, we considered quantum architectures designed for the engineering of the N00N state, a bipartite maximally entangled state crucial in quantum metrology applications.  The fundamental concept underlying these schemes is the transformation of the initial state $|N\rangle
\otimes |0\rangle$ to the N00N state $\frac{1}{\sqrt{2}} (|N\rangle
\otimes|0\rangle +|0\rangle \otimes|N\rangle)$. We showed that this state can be generated as a superposition of modes of quantum light, a combination of light and motion, or a superposition of two spin ensembles.
The approach discussed here can generate mesoscopic and macroscopic entangled states, such as entangled coherent and squeezed states, as well. We show that a large class of  maximally entangled states can be achieved in such an architecture.
The extension of these state engineering methods to the multi-mode setting was also discussed.

\end{document}